\documentclass[reprint,superscriptaddress,amsmath,amssymb,pra,floatfix]{revtex4-2}

\usepackage[utf8]{inputenc}
\usepackage[english]{babel}
\usepackage{amsmath, bm}
\usepackage{amssymb}
\usepackage{placeins}
\usepackage{braket}
\usepackage{booktabs}
\usepackage{physics}
\usepackage{cleveref}
\usepackage{siunitx}
\usepackage{xfrac}
\usepackage{natbib}
\makeatletter

\renewcommand*\env@matrix[1][\arraystretch]{%
  \edef\arraystretch{#1}%
  \hskip -\arraycolsep
  \let\@ifnextchar\new@ifnextchar
  \array{*\c@MaxMatrixCols c}}
\makeatother

\begin{document}

\preprint{APS/123-QED}

\title{Identifying regions of minimal back-scattering by a relativistically-moving sphere}

\author{Mitchell R. Whittam}
\email{mitchell.whittam@kit.edu}
\affiliation{
Institut f\"ur Theoretische Festk\"orperphysik, Karlsruhe Institute of Technology, 76131 Karlsruhe, Germany}%
\author{Aristeidis G. Lamprianidis} 
\affiliation{
Institut f\"ur Theoretische Festk\"orperphysik, Karlsruhe Institute of Technology, 76131 Karlsruhe, Germany}%
\author{Yannick Augenstein}
\affiliation{
Institut f\"ur Theoretische Festk\"orperphysik, Karlsruhe Institute of Technology, 76131 Karlsruhe, Germany}%
\author{Carsten Rockstuhl}
\affiliation{
Institut f\"ur Theoretische Festk\"orperphysik, Karlsruhe Institute of Technology, 76131 Karlsruhe, Germany}%
\affiliation{Institute of Nanotechnology, Karlsruhe Institute of Technology, 76021 Karlsruhe, Germany}

\begin{abstract}
The far-field back-scattering amplitude of an electric field from a relativistically-moving sphere is analyzed.
Contrary to prior research, we do so by expressing the fields in the helicity basis, and we highlight here its advantages when compared to the commonly-considered parity basis.
With the purpose of exploring specific scattering phenomena considering relativistic effects, we identify conditions that minimize the back-scattered field, leading to a relativistic formulation of the first Kerker condition.
The requirements to be satisfied by the sphere are expressed in terms of Mie angles, which constitute an effective parametrization of any possible optical response a sphere might have.
We are able to identify multiple combinations of Mie angles up to octupolar order via gradient-based optimization that satisfy our newly formulated relativistic Kerker condition, yielding minima for the back-scattered energy as low as \SI{0.016}{\percent} of the average scattered energy.
Our results can be extended to involve multiple particles forming a metasurface, potentially having direct implications on the design of light sails as considered by the Breakthrough Starshot Initiative.
\end{abstract}


\maketitle

\section{Introduction}

The scattering of light by a sphere is a canonical problem in optics and electrodynamics and has been investigated for many years, particularly for stationary spheres~\cite{sinclair1947light, kunz1983dynamic, graaff1992reduced, tzarouchis2018light, mundy1974mie, sorensen2000patterns, wriedt2012mie, drake1985mie, geng2004mie, pavlyukh2004nonlinear, hightower1988resonant}.
The scattering of light by spheres is best described using Mie theory, which involves expressing the incident and scattered electromagnetic fields in terms of vector spherical harmonics (VSHs). The amplitude coefficients that weight these VSHs are collected in a vector and are mutually linked by a matrix-vector product.
Moreover, all optical properties of the object are captured by the corresponding matrix, called the transition or T-matrix. For an arbitrary object, the T-matrix can be dense, but is diagonal for a sphere, and the diagonal entries are the Mie coefficients~\cite{mishchenko2002scattering, bohren2008absorption}.

Controlling an object's geometrical and material properties provides a unique way of tailoring the scattered field on demand, and many intriguing aspects have been explored, an example being the so-called Kerker condition~\cite{shamkhi2019transverse, liu2018generalized, poshakinskiy2019optomechanical, babicheva2017resonant, shamkhi2019transparency}.
The first Kerker condition contains the necessary composition of multipolar excitations such that the object exhibits zero back-scattering.
A second Kerker condition implies a vanishing scattering in the forward direction, but this is considered less often, since optical gain is necessary for its observation~\cite{olmos2020kerker}.

While initially formulated for objects that can be safely described in dipolar approximation, it was soon realized that similar effects are encountered while capitalizing on higher-order multipole moments.
This coined the notion of generalized Kerker conditions~\cite{alaee2015generalized}, and Kerker effects have been explored in a large variety of settings. These studies are motivated by high-impact applications related to nanoantennas, chiral molecules, and metamaterials, to name a few~\cite{alaee2015generalized, zhang2020constructive, barhom2019biological, asadchy2022parametric, bukharin2022transverse, zambrana2013duality}.

This paper provides a further yet, to our knowledge, unexplored perspective on the Kerker effect. It considers the Kerker effect in the relativistic regime.
The basic setting of our exploration is that of a relativistically moving sphere illuminated with a monochromatic Gaussian beam characterized by an incident angle $\Theta_{\mathrm{i}}$ relative to the direction of motion of the sphere.
Of course, unlike the case of a stationary sphere, one cannot assume that there exists a combination of multipole excitations that yield zero back-scattering with the inclusion of motion.
However, one can aim to minimize the back-scattering, which depends on the multipolar contribution to the scattering response, for a given speed and incident electric field angle. This leads to an approximate Kerker condition in the case of a relativistically moving sphere.

Our work has clear implications for future technology developments. For example, within the Breakthrough Starshot Initiative~\cite{parkin2018breakthrough, daukantas2017breakthrough}, micro-gram satellites equipped with light sails, potentially made from metasurfaces consisting of a tailored arrangement of scatterers, are to be accelerated with an Earth-based laser system up to a significant fraction of the speed of light.
Using these satellites, neighbouring galaxies shall be explored. The design of such systems has many facets, and among them is the accurate description of the optical response from scattering objects in the form of metasurfaces. The formulation of the light scattering by an isolated object under relativistic conditions, as a pursuit in this contribution, is an important prerequisite to study such more advanced devices.

The structure of the paper is as follows. In Section~\ref{sec:scenario}, the physical setup is outlined, and all necessary coordinate systems are defined.
Moreover, the field of the considered incident beam is transformed from the lab frame to the reference frame of the sphere based on the transformation of each constitutive plane wave of its angular spectrum representation. Afterwards, the scattered field is obtained by solving an ordinary Mie problem in the rest frame of the sphere. We rely on a parametrization of its response using Mie angles~\cite{rahimzadegan2020minimalist}.
These Mie angles constitute a minimalist model to express all possible responses from a sphere, which allows a generic analysis of the back-scattering response. To conclude Section~\ref{sec:scenario}, the scattered field will be transformed back to the lab frame, in which the back-scattering is observed.

In Section~\ref{sec:rel_kerker}, the back-scattering amplitude is visualized with respect to some given Mie angles for a sphere with a fixed velocity and a field with a fixed incident angle.
We implement all calculations using the Julia programming language~\cite{bezanson2017julia} and implement a gradient-based optimization scheme by leveraging automatic differentiation within the JuMP modelling framework~\cite{dunning2017modeling}, much in the spirit of recent works on differentiable physics solvers~\cite{hughes2019forward, minkov2020inverse}.
Using this scheme, we design spheres that provide minimum values for the back-scattering and identify the corresponding combinations of Mie angles. We find multiple suitable combinations, and minimize the back-scattered energy to a negligible \SI{0.016}{\percent} of the average scattered energy. In Section~\ref{sec:conclusion}, we conclude our findings.

\section{Description of the scattering scenario}\label{sec:scenario}

\begin{figure*}[ht]
\centering
\includegraphics[width = 1.0\textwidth]{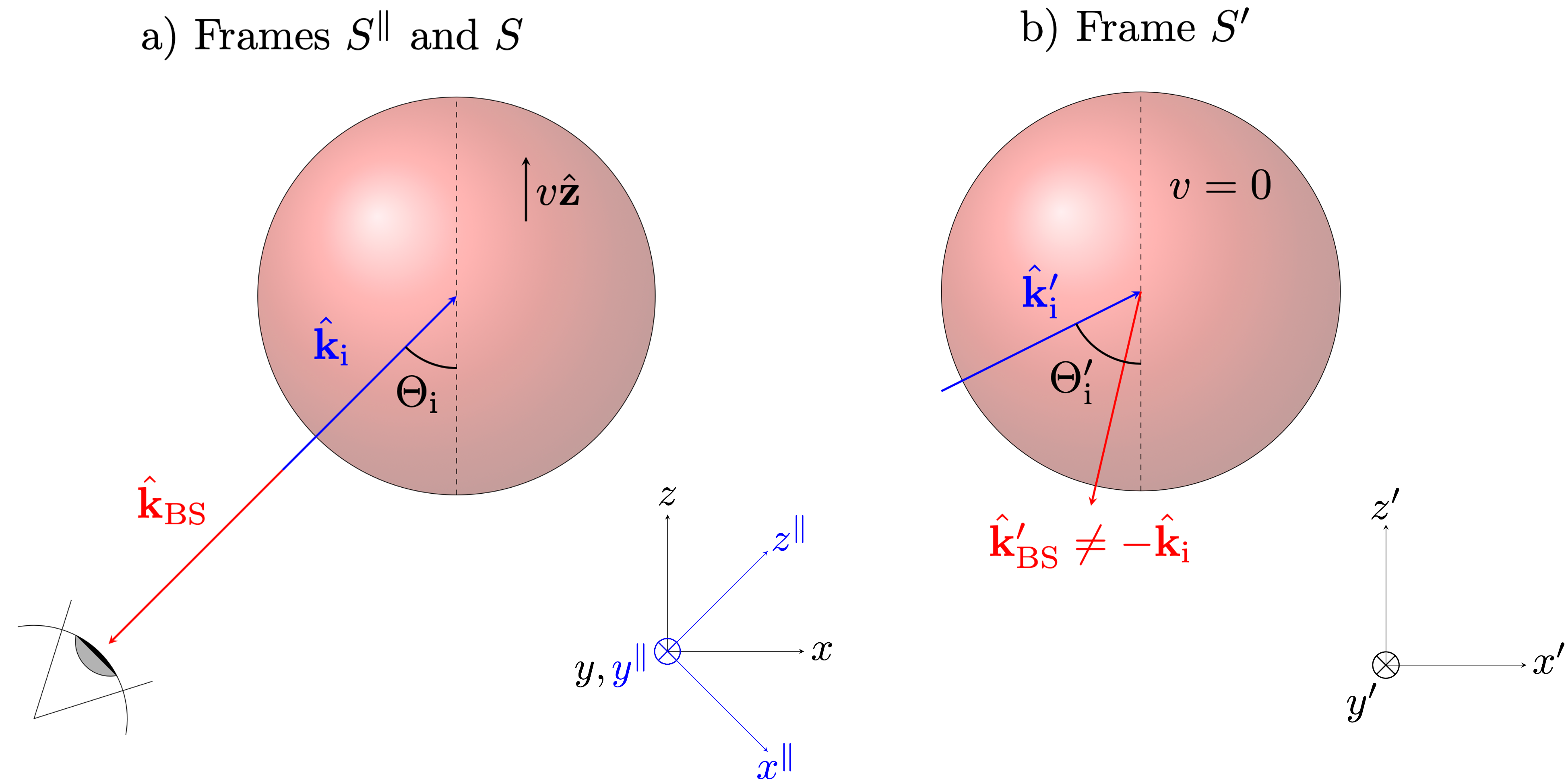}
\caption{\raggedright Pictorial representation of the sphere in a) the beam's frame $S^{\parallel}$ and the rotated frame $S$ containing an external observer (represented by the eye) and b) the sphere's inertial reference frame $S'$. In $S$, the sphere is seen to be moving with velocity $v\hat{\mathbf{z}}$. The wave vector $\hat{\mathbf{k}}_{\mathrm{i}}$ is incident on the sphere with angle $\Theta_{\mathrm{i}}$. In a), the direction of back-scattering $\hat{\mathbf{k}}_{\mathrm{BS}}$ (denoted by the red arrow) is in the opposite direction to $\hat{\mathbf{k}}_{\mathrm{i}}$ and it is this direction that shall be considered when formulating a relativistic Kerker condition. In b), the sphere is stationary ($v=0$) while the Lorentz-boosted wave vector of the incident field is given by $\hat{\mathbf{k}}'_{\mathrm{i}}$ and is incident with angle $\Theta'_{\mathrm{i}}\neq\Theta_{\mathrm{i}}$ when $\Theta_{\mathrm{i}} \notin \{0,\pi\}$, and $\Theta'_{\mathrm{i}}=\Theta_{\mathrm{i}}$ when $\Theta_{\mathrm{i}} \in \{0,\pi\}$. Moreover, the direction of back-scattering $\hat{\mathbf{k}}_{\mathrm{BS}}'$ in this frame is in general not opposite to $\hat{\mathbf{k}}_{\mathrm{i}}$ or $\hat{\mathbf{k}}'_{\mathrm{i}}$.}\label{sphere_both_frames}
\end{figure*}

Before delving into the mathematical description of the scattered field, it is first necessary to specify the geometry and constraints of the system.
We consider a spherical particle moving at a relativistic velocity $\mathbf{v}=v\hat{\mathbf{z}}$ within a pervading incident electric field $\mathbf{E}_{\mathrm{i}}(\mathbf{r}, t)$ with incident angle $\Theta_{\mathrm{i}}$ as observed by an external lab frame $S$.
Although an accompanying magnetic field will always exist, to avoid repetition, we omit explicit reference to this.
Two further frames are considered, namely the beam's frame $S^{\parallel}$, which is the frame where the direction of motion of the beam moves parallel to its corresponding $z^{\parallel}$-axis, and the boosted frame $S'$, which represents the inertial reference frame of the sphere (see Fig.~\ref{sphere_both_frames}).
Accordingly, the corresponding quantities are denoted without a prime in $S$ and with a prime in $S'$, while all quantities in $S^{\parallel}$ are denoted with a $\parallel$ superscript.

A further quantity of interest is the polar angle $\Theta_{\mathrm{i}}$ between $\hat{\mathbf{k}}_{\mathrm{i}}$ and $\mathbf{v}$, i.e., the angle between the beam's propagation direction and the axis of movement of the scatterer (see Fig.~\ref{sphere_both_frames}).
Given the symmetry of the system, we set the azimuthal angle of the incident field $\Phi_{\mathrm{i}}$ to be zero.
Moreover, the direction of back-scattering is given by $\hat{\mathbf{k}}_{\text{BS}}$, the opposite direction to $\hat{\mathbf{k}}_{\mathrm{i}}$.

To determine the scattered field in $S$, we implement the `frame-hopping method' (FHM) as described in~\citet{garner2017time}. For reference, this process is outlined below:
\begin{enumerate}
    \item Lorentz-boost the incident electric field from $S$ to $S'$.
    \item Solve the scattering problem in $S'$.
    \item Inverse Lorentz-boost the scattered field from $S'$ back to $S$. 
\end{enumerate}
The reason for computing the scattered field in $S'$ and not $S$ is a matter of mathematical simplicity. In $S'$, the scattering calculation is analogous to a stationary system, thus avoiding any superfluous variable transformations.

\subsection{Lorentz-boosting the incident field into the scatterer's reference frame}

First, we need to consider the incident field in the beam's reference frame $S^{\parallel}$.
As an incident field, we consider a single monochromatic Gaussian beam of well-defined helicity (\emph{i.e.}, handedness) expanded in terms of circularly polarized plane waves, which are eigenstates of the electromagnetic wave equation. We use the following ket in abstract Dirac notation to denote such plane waves as eigenstates of free space characterized by helicity $\lambda^{\parallel}=\pm1$, temporal frequency $\omega^{\parallel}$, and direction of propagation $\hat{\mathbf{k}}^{\parallel}$:
\begin{equation}\label{plane_wave_def}
\ket{\lambda^{\parallel}\ \hat{\mathbf{k}}^{\parallel}\ \omega^{\parallel}} \doteq \hat{\mathbf{e}}_{\lambda}(\hat{\mathbf{k}}^{\parallel})\exp\{{\mathrm{i}\omega^{\parallel}[(\hat{\mathbf{k}}^{\parallel}\cdot\mathbf{r}/c) - t}]\} \qq{,}
\end{equation}
where the symbol $\doteq$ refers to the spatiotemporal representation of the plane wave eigenstate. The quantity $c$ is the speed of light in vacuum, and the polarization unit vector $\hat{\mathbf{e}}_{\lambda^{\parallel}}(\hat{\mathbf{k}}^{\parallel})$ is given by
\begin{align}\label{polarisation_unit_vec}
   \hat{\mathbf{e}}_{\lambda^{\parallel}}(\hat{\mathbf{k}}^{\parallel}) &= \frac{-\lambda^{\parallel}\hat{\theta}(\hat{\mathbf{k}}^{\parallel})-\mathrm{i}\hat{\phi}(\hat{\mathbf{k}}^{\parallel})}{\sqrt{2}} \qq{,}
\end{align}
where $\lambda^{\parallel} = \pm 1$ corresponds to left/right circularly-polarized waves. The quantities $\hat{\theta}$ and $\hat{\phi}$ are, respectively, the polar and azimuthal spherical unit vectors perpendicular to the direction of propagation $\hat{\mathbf{k}}^{\parallel}$ that is characterized by the polar and azimuthal angles of propagation $\theta^{\parallel}$, $\phi^{\parallel}$ of each constituent plane wave.

Denoting quantities that belong to the incident field with the subscript `$\mathrm{i}$', we represent a general electric field in terms of its angular spectrum, i.e., as a plane wave expansion:
\begin{align}\label{E_inc_lab}
\ket{\mathbf{E}^{\parallel}_{\mathrm{i}}}
&=
\sum_\lambda\int_{0}^{2\pi}\mathrm{d}\phi^{\parallel}\int^{\pi}_{0}\mathrm{d}\theta^{\parallel}\int^{\infty}_{0^{+}}\mathrm{d}\omega^{\parallel}\nonumber\\
&
\hspace{2pt}\mathcal{G}^{\parallel}_{\lambda^{\parallel},\hspace{.4mm}\mathrm{i}}(\omega^{\parallel},\theta^{\parallel},\phi^{\parallel})\ket{\lambda^{\parallel}\ \hat{\mathbf{k}}^{\parallel}\ \omega^{\parallel}}  + \mathrm{c.c.} \qq{,}
\end{align}
where the amplitudes for a monochromatic Gaussian beam focused at the origin of $S$ of waist $w_0$ , frequency $\omega_\mathrm{i}$ and helicity $\lambda_\mathrm{i}$ propagating along the $+z$-axis are given by:
\begin{align}\label{gauss_profile}
\mathcal{G}^{\parallel}_{\lambda^{\parallel},\hspace{.4mm}\mathrm{i}}(\omega^{\parallel},\theta^{\parallel},\phi^{\parallel})
&=
E_{0}\sin(2\theta^{\parallel})\nonumber\\
&\cdot
\exp[-\frac{\omega_\mathrm{i}^2 w_{0}^{2}\sin^2(\theta^{\parallel})}{4c^2}]\nonumber\\
&\cdot
\delta_{\lambda^{\parallel}\lambda_\mathrm{i}}\delta(\omega^{\parallel}-\omega_\mathrm{i})H(\pi/2-\theta^{\parallel}) \qq{,}
\end{align} 
where $E_{0}$ is a constant, and $H(\pi/2 - \theta^{\parallel})$ is the Heaviside step function which eliminates all counter-propagating waves.
Moreover, we consider the waist $w_0$ to be very large, such that $\ket{\mathbf{E}^{\parallel}_{\mathrm{i}}}$ approximates to a plane wave but is nonetheless still finite in space.
The reason for doing this is that, for a regular plane wave infinitely extended in space, the interaction of the incident wave with the moving sphere would be incessant and, therefore, the scattered power flux would have a cylindrical symmetry with respect to the axis of movement of the scatterer. That is, the scattered power flux would be translationally invariant with respect to this axis and would only vary azimuthally. On the other hand, an excitation of finite spatial extent ensures that the interaction of light with the scatterer is localized in space (around the origin of $S$), therefore, yielding a spherical-like scattering of waves emanating from the region where the interaction takes place.

To consider an electric field of arbitrary angle of incidence $\Theta_{\mathrm{i}}$, we apply a rotation operator $\hat{\mathbf{R}}_y(\Theta_{\mathrm{i}})$ about the $y$-axis to Eqn.~(\ref{E_inc_lab}) to transit from a representation of the beam with respect to the $S^{\parallel}$ to one with respect to $S$ such that
\begin{align}\label{E_to_ER}
     \ket{\mathbf{E}_{\mathrm{i}}} = \hat{\mathbf{R}}_y(\Theta_{\mathrm{i}})\ket{\mathbf{E}^{\parallel}_{\mathrm{i}}} \qq{,}
\end{align}
where 
\begin{align}\label{rotate_PW}
    \hat{\mathbf{R}}_y(\Theta_{\mathrm{i}})\ket{\lambda^{\parallel}\ \hat{\mathbf{k}}^{\parallel}\ \omega^{\parallel}} 
    &=
    \sum_{\lambda}\int_{0}^{2\pi}\mathrm{d}\phi\int^{\pi}_{0}\mathrm{d}\theta\int^{\infty}_{0^{+}}\mathrm{d}\omega\nonumber\\
    &
    \hspace{2pt}\mathcal{R}_{\lambda^{\parallel},\lambda}(\omega^{\parallel},\theta^{\parallel},\phi^{\parallel},\omega,\theta,\phi;\Theta_{\mathrm{i}})\nonumber\\
    &
    \ket{\lambda\ \hat{\mathbf{k}}\ \omega} \qq{.}
\end{align}
The transformation coefficients are given by
\begin{align}\label{rotate_ampl}
    \mathcal{R}_{\lambda^{\parallel},\lambda}(\omega^{\parallel},\theta^{\parallel},\phi^{\parallel},\omega,\theta,\phi;\Theta_{\mathrm{i}})
    &=
    \mathcal{P}(\theta^{\parallel}, \phi^{\parallel},\Theta_{\mathrm{i}})\nonumber\\
    &\cdot
    \delta_{\lambda\lambda^{\parallel}}\nonumber\\
    &\cdot
    \delta(\omega-\omega^{\parallel})\nonumber\\
    &\cdot
    \delta\left[\theta-\mathrm{arccos}(\hat{k}_{z})\right]\nonumber\\
    &\cdot
    \delta\left[\phi-\mathrm{atan2}(\hat{k}_{y},\hat{k}_{x})\right] \qq{,}
\end{align}
where
\begin{align}
    \begin{pmatrix}[1.3]
    \hat{k}_{x}\\
    \hat{k}_{y}\\
    \hat{k}_{z}
    \end{pmatrix}
    &=
    \begin{pmatrix}[1.3]
    \cos{\Theta_{\mathrm{i}}}&0&\sin{\Theta_{\mathrm{i}}}\\
    0&1&0\\
    -\sin{\Theta_{\mathrm{i}}}&0&\cos{\Theta_{\mathrm{i}}}
    \end{pmatrix}    
    \begin{pmatrix}[1.3]
    \hat{k}^{\parallel}_{x}\\
    \hat{k}^{\parallel}_{y}\\
    \hat{k}^{\parallel}_{z}
    \end{pmatrix}\nonumber\\
    &=
    \begin{pmatrix}[1.3]
    \sin\theta^{\parallel}\cos\phi^{\parallel}\cos{\Theta_{\mathrm{i}}}+\cos\theta^{\parallel}\sin{\Theta_{\mathrm{i}}}\\
    \sin\theta^{\parallel}\sin\phi^{\parallel}\\
    -\sin\theta^{\parallel}\cos\phi^{\parallel}\sin{\Theta_{\mathrm{i}}}+\cos\theta^{\parallel}\cos{\Theta_{\mathrm{i}}}
    \end{pmatrix} \qq{,}
\end{align}
and $\mathcal{P}(\theta^{\parallel}, \phi^{\parallel},\Theta_{\mathrm{i}})$ is a prefactor corresponding to the acquired phase due to the rotation:
\begin{align}\label{phase_expon}
    \mathcal{P}(\theta^{\parallel}, \phi^{\parallel},\Theta_{\mathrm{i}}) =
    \mathrm{exp}[\mathrm{i}p(\theta^{\parallel}, \phi^{\parallel},\Theta_{\mathrm{i}})] \qq{,}
\end{align}
with
\begin{align}\label{phase angle}
    p(\theta^{\parallel}, \phi^{\parallel},\Theta_{\mathrm{i}}) 
    &= 
    \mathrm{atan2}\Big[- \lambda  \sin (\Theta_{\mathrm{i}}) \sin (\phi^{\parallel}),\nonumber\\
    &\hspace{0.41cm}\hspace{0.1cm}\cos (\theta^{\parallel} ) \sin (\Theta_{\mathrm{i}} ) \cos (\phi^{\parallel} )\nonumber\\
    &
    \hspace{0.05cm}+\sin (\theta^{\parallel} ) \cos (\Theta_{\mathrm{i}} )\Big] \qq{.}
\end{align}

After doing this, one can follow the first step of the frame-hopping method and compute the Lorentz boost of the incident electric field from $S$ to $S'$. In App.~\ref{LB_of_plane_waves}, we calculate the Lorentz boost of plane waves.
We denote with $\hat{\mathbf{L B}}_z(\beta)$ the operator that boosts fields along the $z$-axis with speed $v=\beta c$, where $0\leq\beta<1$ from $S$ to $S'$.
This operator acts on the eigenstates of monochromatic plane waves with well-defined helicity in the following way:
\begin{align}\label{projection}
\hat{\mathbf{LB}}_{z}(\beta)\ket{\lambda\ \hat{\mathbf{k}}\ \omega} 
&=
\sum_{\lambda'}\int_{0}^{2\pi}\mathrm{d}\phi'\int^{\pi}_{0}\mathrm{d}\theta'\int^{\infty}_{0^{+}}\mathrm{d}\omega'\nonumber\\
&
\hspace{0.5cm}\mathcal{L}_{\lambda\lambda'}(\omega,\theta,\phi,\omega',\theta',\phi';\beta)\nonumber\\
&\hspace{0.4cm}\ket{\lambda'\ \hat{\mathbf{k}}'\ \omega'} \qq{,}
\end{align}
with the transformation coefficients given by
\begin{align}\label{boost_coefficients}
    \mathcal{L}_{\lambda\lambda'}(\omega,\theta,\phi,\omega',\theta',\phi';\beta)
    &=
    \mathcal{C}\left(\beta,\theta\right)\nonumber\\
    &\cdot
    \delta_{\lambda'\lambda}\nonumber\\
    &\cdot
    \delta\left(\omega'-\mathcal{C}\left(\beta,\theta\right)\omega\right)\nonumber\\
    &\cdot
    \delta\left[\theta'-\mathrm{arccos}\left(\frac{\cos\theta-\beta}{1-\beta\cos\theta}\right)\right]\nonumber\\
    &\cdot
    \delta\left(\phi'-\phi\right) \qq{,}
\end{align}
where $\gamma=1/\sqrt{1-\beta^2}$, $\cos\theta=\hat{\mathbf{k}}\cdot\hat{\mathbf{z}}$ and 
\begin{align}\label{C}
\mathcal{C}\left(\beta,\theta\right)=\gamma\left[1-\beta\cos\theta\right] \qq{,}
\end{align}
which is derived in App.~\ref{LB_of_plane_waves}.
We see from Eqn.~(\ref{boost_coefficients}) that
\begin{align}\label{theta_boost}
    \theta'=\mathrm{arccos}\left(\frac{\cos\theta-\beta}{1-\beta\cos\theta}\right)\qq{,}
\end{align}
and
\begin{align}\label{freq_boost}
    \omega'=\mathcal{C}\left(\beta,\theta\right)\omega
\end{align}
correspond to the Lorentz boost of $\theta$ and $\omega$, respectively. Since the motion occurs solely along the $z$-axis, the azimuthal angle $\phi$ remains unchanged under the Lorentz boost, that is,
\begin{align}\label{phi_boost}
    \phi'=\phi \qq{.}
\end{align}

\begin{figure}[ht]
\centering
\includegraphics{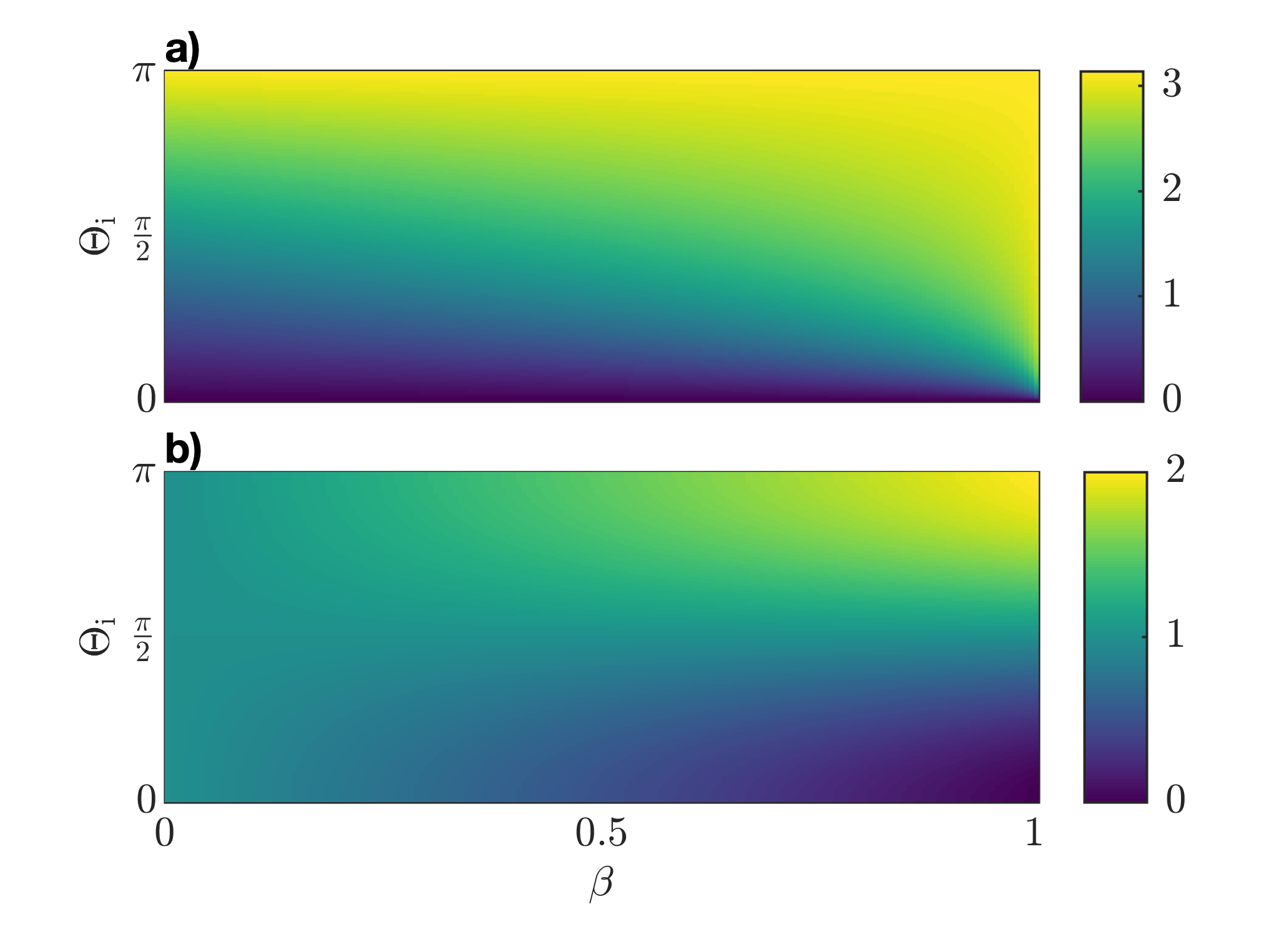}
\caption{\raggedright a) The Doppler shift $\Theta'_{\mathrm{i}}$ of $\Theta_{\mathrm{i}}$ as a function of $\Theta_{\mathrm{i}}$ and $\beta$ determined using Eqn.~\eqref{theta_boost}. This demonstrates that the direction of the incident wave as perceived in $S'$ is different to that in $S$.
b) The normalized Doppler-shifted incident frequency $\omega'_{\mathrm{i}}/\gamma\omega_{\mathrm{i}}$ as a function of $\Theta_{\mathrm{i}}$ and $\beta$, where $\omega'_{\mathrm{i}}$ is the Doppler-shifted incident frequency determined using Eqn.~(\ref{freq_boost}). When $\Theta_{\mathrm{i}}=0$ and $\beta\to 1$, the object is moving away from the field source, thus causing the incident frequency in $S'$ to decrease (redshift). When $\Theta_{\mathrm{i}}=\pi$ and $\beta\to 1$, the object is moving towards the field source, thus causing the incident frequency in $S'$ to increase (blueshift). When $\Theta_{\mathrm{i}}=\pi/2$, $\omega'_{\mathrm{i}}/\gamma\omega_{\mathrm{i}} = 1$ for all $\beta$. This is due to $\hat{\mathbf{k}}_{\mathrm{i}}$ and $\mathbf{v}$ being perpendicular to each other. The factor $1/\gamma$ is necessary to eliminate the over-exaggeration of $\omega'_{\mathrm{i}}$ when $\beta\to 1$ and $\Theta_{\mathrm{i}}\to\pi$. Without this factor, the other frequency values would appear too suppressed.}
\label{freq_ratio}
\end{figure}

The Lorentz boost $\theta'$ of $\theta$ given by Eqn.~(\ref{theta_boost}) explains the perceived change in direction of the beam in $S'$ compared to $S$ as shown by $\hat{\mathbf{k}}_{\mathrm{i}}$ and $\hat{\mathbf{k}}_{\mathrm{i}}'$ in Fig.~\ref{sphere_both_frames}.
In Fig.~\ref{freq_ratio}~a), this is visualized for the Lorentz boost $\Theta'_{\mathrm{i}}$ of the polar angle of the incident field with respect to the incident angle $\Theta_{\mathrm{i}}$ as seen in $S$ and speed ratio $\beta$. Moreover, the Doppler shift $\omega'$ of $\omega$ is displayed in Fig.~\ref{freq_ratio}~b) with the same functional dependency. Note that, for an incident angle of $\Theta_{\mathrm{i}}=0$ and a speed ratio $\beta\to1$, the Doppler-shifted frequency becomes zero. This corresponds to the sphere moving away from the external observer in $S$ at a speed tending to that of light, thus exhibiting a complete redshift. In other words, the incident wave is perceived by the sphere to be so stretched out that the frequency disappears in its reference frame. Conversely, when $\Theta_{\mathrm{i}}=\pi$ and $\beta\to1$, the wave is seen to be infinitely blueshifted in $S'$, corresponding to a completely compressed wave with infinite frequency. This corresponds to the sphere moving towards the source of the incident field.

Note that the same expression for the scaling factor $\mathcal{C}\left(\beta,\theta\right)$ is given by Eqn.~(27) in~\citet{de2000electromagnetic}.
Importantly, we observe that for Eqn.~(\ref{boost_coefficients}) to be non-zero, the helicity of the field must remain invariant upon the Lorentz boost transformation due to the $\delta_{\lambda'\lambda}$ term.
This invariance demonstrates the power of expressing the fields in the helicity instead of the parity basis, that is, specifically making use of circularly-polarized plane waves instead of TE/TM plane waves.

Generally speaking, the change in direction of the beam upon boosting is given by the following transformation of the wavevectors:
\begin{align}\label{LB_k}
   \hat{\mathbf{k}}' = \frac{\hat{\mathbf{k}}+\left[(\gamma-1)\cos\theta-\gamma\beta\right]\hat{\mathbf{z}}}{\mathcal{C}\left(\beta,\theta\right)} \qq{.}
\end{align}

Finally, putting all the above together, and after some straightforward algebra, we can get the following relation between the amplitudes of the initially considered and non-rotated incident beam in $S^{\parallel}$ and the rotated one in $S'$:
\begin{align}\label{E_inc_object_final} 
    \ket{\mathbf{E}_{\mathrm{i}}'} 
    &=
    \hat{\mathbf{LB}}_z(\beta)\hat{\mathbf{R}}_y(\Theta_{\mathrm{i}})\ket{\mathbf{E}_{\mathrm{i}}^{\parallel}}\nonumber\\ 
    &= 
    \sum_{\lambda'}\int_{0}^{2\pi}\mathrm{d}\phi'\int^{\pi}_{0}\mathrm{d}\theta'\int^{\infty}_{0^{+}}\mathrm{d}\omega'\nonumber\\
    &
    \hspace{0.5cm}\mathcal{G}'_{\lambda',\hspace{.4mm}\mathrm{i}}(\omega',\theta',\phi')\ket{\lambda_{\mathrm{i}}'\ \hat{\mathbf{k}}_{\mathrm{i}}'\ \omega_{\mathrm{i}}'}\nonumber\\
    &+
    \hspace{0.08cm}\mathrm{c.c.} \qq{,}
\end{align}
where 
$\omega'_{\mathrm{i}}$ and $\hat{\mathbf{k}}'_{\mathrm{i}}$ are determined by Eqns.~(\ref{freq_boost}) and (\ref{LB_k}), respectively and
\begin{align}\label{G_S_prime}
\mathcal{G}'_{\lambda',\hspace{.4mm}\mathrm{i}}(\omega',\theta',\phi')
&=
\mathcal{J}(\theta',\phi',\Theta_{\mathrm{i}})\nonumber\\
&\cdot
\mathcal{G}^{\parallel}_{\lambda',\hspace{.4mm}\mathrm{i}}\Bigg\{\frac{\omega'}{\mathcal{C}\left[\beta, \theta^{\parallel}(\theta',\phi')\right]},\nonumber\\
&
\theta^{\parallel}(\theta',\phi'),\phi^{\parallel}(\theta',\phi')\Bigg\} \qq{,}
\end{align}
where $\theta^{\parallel}(\theta',\phi')$ and $\phi^{\parallel}(\theta',\phi')$ express $\theta^{\parallel}$ and $\phi^{\parallel}$ as viewed from $S'$:
    \begin{align}\label{theta_alt}
    \theta^{\parallel}(\theta',\phi') 
    &=
    \arccos\Bigg\{\frac{1}{{\gamma(1 + \beta\cos\theta')}}\nonumber\\
    &\cdot
    \Big[\sin\theta'\cos\phi'\sin\Theta\nonumber\\
    &+
    \gamma(\cos\theta' + \beta)\cos\Theta\Big]\Bigg\},\\
    \label{phi_alt}
    \phi^{\parallel}(\theta',\phi') 
    &= \mathrm{atan2}[\sin\theta'\sin\phi',\nonumber\\
    &\hspace{0.4cm}\sin\theta'\cos\phi'\cos\Theta\nonumber\\
    &- \gamma(\cos\theta' + \beta)\sin\Theta] \qq{,}
\end{align}
with the Jacobian
\begin{align}\label{jacobian}
\mathcal{J}(\theta',\phi',\Theta_{\mathrm{i}})
&=
\begin{vmatrix}[1.8]
    \frac{\partial\theta^{\parallel}}{\partial\theta'} & \frac{\partial\theta^{\parallel}}{\partial\phi'}\\
    \frac{\partial\phi^{\parallel}}{\partial\theta'} & \frac{\partial\phi^{\parallel}}{\partial\phi'}
    \end{vmatrix}
\end{align}
that converts $\mathrm{d}\phi^{\parallel}\mathrm{d}\theta^{\parallel}$ to $\mathrm{d}\phi'\mathrm{d}\theta'$.

Recall that we do this since, for simplicity, we wish to carry out the scattering calculation in $S'$, that is, where it is equivalent to that in the stationary case.
For a given incident field described by Eqn.~({\ref{gauss_profile}}), we can use Eqn.~(\ref{G_S_prime}) to calculate the amplitudes that are needed to describe the incident field in $S'$ in terms of the plane wave representation given by Eqn.~(\ref{E_inc_object_final}).

\subsection{Solving the scattering problem in the sphere's reference frame}

The approach taken to calculate the amplitude of the scattered field begins by expressing the incident field $\ket{\mathbf{E}_{\mathrm{i}}'}$ as a series of spherical waves with respect to the coordinates describing $S'$~\cite{mishchenko2002scattering}:
\begin{equation}\label{E_inc_VSH}
\ket{\mathbf{E}_{\mathrm{i}}'} = \sum_{\lambda'\ell'm'}\int^{\infty}_{0^{+}}\mathrm{d}\omega'\mathcal{A'}_{ \lambda' \ell'm'}(\omega')\ket{\omega'\hspace{2pt} \lambda'\hspace{2pt}\ell'\hspace{2pt}m'}^{(1)}+\mathrm{c.c.}\qq{,}
\end{equation}
where $\ket{\omega'\hspace{2pt} \lambda'\hspace{2pt}\ell'\hspace{2pt}m'}^{(1)}$ signifies a regular VSH attached to $S'$ with frequency $\omega'$, helicity $\lambda'$, multipolar index $\ell'$ ($\ell'=1$ corresponds to dipoles, $\ell'=2$ corresponds to quadrupoles etc.), and angular momentum along the $z$-axis $m'=-\ell',-(\ell'-1),\hspace{1mm}...\hspace{1mm},\ell'$.
The ${(1)}$ superscript denotes that the VSH corresponds to a first-order spherical Bessel function $j_{\ell'}(k'r')$, and the coefficients of the expansion are given by
\begin{align}\label{A_unsimplified}
    \mathcal{A'}_{ \lambda' \ell'm'}(\omega')
    &= 
    \int_{0}^{2\pi}\mathrm{d}\phi'\int^{\pi}_{0}\mathrm{d}\theta'\hspace{2pt}\mathcal{G}'_{\lambda',\hspace{.4mm} \mathrm{i}}(\omega',\theta',\phi')\nonumber\\
    &\cdot
    \mathcal{S}_{\lambda'\ell'm'}(\omega',\hat{\mathbf{k}}')\qq{,}
\end{align}
where the transformation coefficients between the plane waves and the spherical waves (under which transformation the helicity and frequency of the waves remain unchanged) are given by:
\begin{align}\label{spher_expans_of_PW}
    \mathcal{S}_{\lambda'\ell'm'}(\omega',\hat{\mathbf{k}}')
     &=
     \hspace{1pt}4\pi\mathrm{i}^{\ell'+2m'+1}\hspace{1pt} \Omega_{\ell'm'}\nonumber\\
     &\cdot
     \tau^{(\lambda')}_{\ell'm'}[\mathrm{\theta}'(\hat{\mathbf{k}}')] \hspace{1pt}\mathrm{e}^{-  \mathrm{i} m' \phi'(\hat{\mathbf{k}}')} \qq{,}
\end{align}
where $\Omega_{\ell'm'}$ is a normalization constant and $\tau^{(\lambda')}_{\ell'm'}[\mathrm{\theta}'(\hat{\mathbf{k}}')]$ is a function which we define in App.~\ref{VSH_definitions}.
The expression given in Eqn.~(\ref{spher_expans_of_PW}) is derived by applying Eqn.~(\ref{polarisation_unit_vec}) to Eqn.~(12) in~\citet{lamprianidis2018excitation}.

For the case of monochromatic excitation (in $S$), like the one we consider here, we have that
\begin{align}\label{gauss_profile0}
\mathcal{G}^{\parallel}_{\lambda^{\parallel},\hspace{.4mm}\mathrm{i}}(\omega^{\parallel},\theta^{\parallel},\phi^{\parallel})=\mathcal{G}^{\parallel,0}_{\lambda^{\parallel},\hspace{.4mm}\mathrm{i}}(\theta^{\parallel},\phi^{\parallel})\delta(\omega^{\parallel}-\omega_\mathrm{i}) \qq{.}
\end{align} 
Using this expression, we get the following simplified expression for the incident spherical amplitudes in $S'$:
\begin{align}\label{A_simplified}
     \mathcal{A'}_{ \lambda' \ell'm'}(\omega')&=
     -4\pi  i ^{\ell'+2m'+1}\hspace{1pt} \Omega_{\ell'm'}\tau^{(\lambda')}_{\ell'm'}(\theta'_{0}) \nonumber\\
    &\cdot
    \delta\left(\omega'\in\left[\frac{\omega_\mathrm{i}}{\gamma(1+\beta)},\frac{\omega_\mathrm{i}}{\gamma(1-\beta)}\right]\right)\nonumber\\
    &\cdot
    \frac{(1+\beta\cos{\theta'_{0}})}{\beta\sin{\theta'_{0}}}\nonumber\\
    &\cdot
    \int_{0}^{2\pi}\mathrm{d}\phi'\hspace{1pt}\mathrm{e}^{-  \mathrm{i} m' \phi' }\nonumber\\
    &\cdot
    \mathcal{P}\left[\theta^\parallel(\theta'_0,\phi'),\phi^\parallel(\theta'_0,\phi'),\Theta_{\mathrm{i}}\right]\nonumber\\
    &\cdot
    \mathcal{J}(\theta'_{0},\phi',\Theta_{\mathrm{i}})\mathcal{G}^{\parallel,0}_{\lambda',\hspace{.4mm}\mathrm{i}}[\theta^{\parallel}(\theta'_{0},\phi'),\phi^{\parallel}] \qq{,}
\end{align}
where:
\begin{align}\label{roots_f_theta}
    \theta'_{0} = \arccos{\left(\frac{\gamma\omega_{\mathrm{i}} - \beta^{2}\gamma\omega_{\mathrm{i}}-\omega'}{\beta\omega'}\right)} \qq{.}
\end{align}
Next, in conjunction with Step 2 of the FHM, we need to express the scattered field $\ket{\mathbf{E}_\mathrm{s}'}$ in a series of radiating VSHs in $S'$, denoted as $\ket{\omega'\hspace{2pt} \lambda'\hspace{2pt}\ell'\hspace{2pt}m'}^{(3)}$.
Analogous to Eqn.~(\ref{E_inc_VSH}), this can be written as
\begin{equation}\label{E_sca_VSH}
\ket{\mathbf{E}_\mathrm{s}'} = \sum_{\lambda'\ell'm'}\int^{\infty}_{0^{+}}\mathrm{d}\omega'\mathcal{B}'_{\lambda'\ell'm'}(\omega')\ket{\omega'\hspace{2pt} \lambda'\hspace{2pt}\ell'\hspace{2pt}m'}^{(3)}+\mathrm{c.c.}\qq{,}
\end{equation}
where the ${(3)}$ superscript denotes that the VSHs correspond to a third-order spherical Bessel (Hankel) function $h_{\ell'}(k'r')$.
Specific expressions of the radiating and regular VSHs are given in App.~\ref{VSH_definitions}. Moreover, the `s' subscript denotes quantities which correspond to the scattered field.

Finally, the scattering coefficients $\mathcal{B'}_{ \lambda' \ell'm'}(\omega')$ can be related to the incident coefficients $\mathcal{A'}_{ \lambda' \ell'm'}(\omega')$ by way of the T-matrix formalism~\cite{mishchenko2002scattering}:
\begin{equation}\label{t_mat_relation}
    \mathbf{B}' = \mathbf{T}^{\text{H}}\mathbf{A}' \qq{,}
\end{equation}
where $\mathbf{A}'$ and $\mathbf{B}'$ are vectors containing the incident and scattering coefficients, respectively, and $\mathbf{T}^{\text{H}}$ is the corresponding T-matrix expressed in the helicity basis (see App.~\ref{T-matrix_helicity}).
The T-matrix fully describes the scattering response of the individual scatterer in the stationary case, which we can safely use in the rest frame of the sphere.
Let us note that the time-invariance of the stationary system implies a matrix that is diagonal with respect to frequency $\omega'$, whereas duality-symmetry implies a diagonal matrix with respect to helicity $\lambda'$, and the spherical symmetry of the scatterer implies a diagonal matrix with respect to the multipolar indices $\ell', m'$.

Specifically, for a spherical scatterer, we can write

\begin{align}\label{component_B}
\mathcal{B}_{\lambda'\ell'm'}(\omega')=\sum_{\lambda_0}\mathrm{T}_{\lambda'\lambda_0,\ell'}(\omega')\mathcal{A}_{\lambda_0\ell'm'}(\omega')\qq{,}
\end{align}
where $\lambda_{0}=\pm1$ is a dummy index representing helicity and the term $\mathrm{T}_{\lambda'\lambda_0,\ell'}$ is defined at the end of App.~\ref{T-matrix_helicity}.
Moreover, in this work, we will make the assumption that the T-matrix of the scatterer is non-dispersive, \emph{i.e.}, invariant with respect to frequency.
This assumption is logical as long as we are exciting with a monochromatic beam with a narrow angular spectrum, \emph{i.e.}, a large waist. One must consider this, since the plane-wave components of the beam all Doppler-shift differently depending on their polar angles of propagation. However, a small angular width in $S$ minimizes this difference, thus allowing us to assume a non-dispersive T-matrix in $S'$.
As we will see, this assumption significantly simplifies the final equations used for numerical computation.

Finally, we require an expression for the electric field in the far-field region of $S'$.
For this, we need to use the following asymptotic expression for the radiating spherical waves:
\begin{align}
    \lim_{\omega' r' /c\rightarrow\infty}\ket{\omega'\hspace{2pt}\lambda'\hspace{2pt}m'\hspace{2pt}\ell'}^{(3)}
    &\equiv
    (-\mathrm{i} )^{\ell'}\hspace{3pt}\mathbf{f}_{\lambda',\ell' m'}(\hat{\mathbf{r}}')\hspace{3pt}\nonumber\\
    &\cdot
    \frac{\mathrm{e}^{\mathrm{i}\omega'( r' /c-t')}}{\omega' r' /c} \qq{,}
\end{align}
which, from Eqn.~(\ref{E_sca_VSH}), readily gives the following expression for the electric field in the far-field region of $S'$:
\begin{align}\label{E_sca_VSH_ff}
\mathbf{E}^{'\mathrm{ff}}_{\mathrm{s}}(\mathbf{r}', t') 
&=
\sum_{\lambda'\ell'm'}\int^{+\infty}_{0^+}\mathrm{d}\omega'\mathcal{B'}_{\lambda'\ell'm'}(\omega')(-\mathrm{i} )^{\ell'}\nonumber\\
&\cdot
\mathbf{f}_{\lambda',\ell'm'}(\hat{\mathbf{r}}')\hspace{3pt}\frac{\mathrm{e}^{\mathrm{i}\omega'( r' /c-t')}}{\omega' r' /c} +\mathrm{c.c.} \qq{,}
\end{align}
where $\mathbf{f}_{\lambda',\ell' m'}(\hat{\mathbf{r}}')$ is a vector function defined in App.~\ref{VSH_definitions}.

As shown in~\citet{garner2017lorentz}, the angular density of the total radiation energy flux in a given direction in $S'$ specified by $\theta'$ and $\phi'$, which we denote as $U'(\theta', \phi')$, is calculated by integrating the amplitude of the electric field $\mathbf{E}'^{\mathrm{ff}}_\mathrm{s}(\mathbf{r}', t')$ in the far-field limit.
As a result, we have
\begin{align}\label{ang_pow_integral_sphere_frame}
U'(\theta', \phi') = \lim_{r'\to\infty}\int^{\infty}_{-\infty}(r')^{2}\frac{|\mathbf{E}'^{\mathrm{ff}}_{\mathrm{s}}(\mathbf{r}', t')|^{2}}{\eta_{0}}\mathrm{d}t' \qq{,}
\end{align}
where $\eta_{0}$ is the impedance of free space.
An expanded, and numerically-efficient form of Eqn.~(\ref{ang_pow_integral_sphere_frame}) is given in App.~\ref{expansion_U}.

At this point, the second step of the FHM is complete.

\subsection{Solution to the scattering problem in the lab frame}

To investigate the back-scattering, we analyse the directivity $D(\theta, \phi)$ of the sphere.
This is defined as~\cite{poljak2019human}
\begin{align}\label{directivity_general}
    D(\theta, \phi) = \frac{U(\theta, \phi)}{W_{\mathrm{tot}}/4\pi} \qq{,}
\end{align}
where $U(\theta, \phi) = \sum_{\lambda_{\mathrm{s}}}U_{\lambda_{\mathrm{s}}}(\theta, \phi)$ is the angular density of the total radiation energy in a given direction in $S$ specified by $\theta$ and $\phi$, $U_{\lambda_{\mathrm{s}}}$ is the component of $U(\theta, \phi)$ corresponding to the scattered helicity $\lambda_{\mathrm{s}}=\pm1$, and $W_{\mathrm{tot}}$ is the total scattered energy.

Considering the directivity of the sphere allows us to obtain a physically meaningful and intuitive formulation from which the behaviour of the back-scattering can be interpreted.
Qualitatively speaking, the directivity is the ratio of the total angular energy $U(\theta, \phi)$ to the average scattered energy $W_\mathrm{tot}/4\pi$ by an analogous isotropic scatterer.
Consequently, a directivity $>$1 means that the contribution of back-scattered energy outweighs that of the average scattered energy.
Conversely, a directivity  $<$1 implies that the back-scattered energy is lower than the average energy scattered by the sphere.

We are now in a position to carry out the final step of the FHM, that is, transforming the directivity from $S'$ back to $S$. The power of the FHM really becomes apparent here, since the angular energy $U(\theta, \phi)$ in $S$ can easily be related to quantities in $S'$.
More specifically, we have
\begin{align}\label{tot_scat_pow_simplified_lab}
W_{\mathrm{tot}}
&=
\int^{2\pi}_{0}\int^{\pi}_{0}U(\theta, \phi)\sin{\theta} \mathrm{d}\theta\mathrm{d}\phi \qq{,}
\end{align}
where, as we see from Eqn.~(21) in~\citet{garner2017lorentz},
\begin{align}\label{energy_bs_lab}
U(\theta, \phi) = [\gamma\left(1+\beta{\cos\theta'}\right)]^{3}\hspace{0.1cm}U'(\theta', \phi') \qq{,}
\end{align}
where $\theta'$ and $\phi'$ can be transformed using Eqns.~(\ref{theta_boost}) and~(\ref{phi_boost}) to obtain an expression for $U(\theta, \phi)$ in $S$.
For back-scattering, we have $\theta=\Theta_{\mathrm{BS}}$ and $\phi=\Phi_{\mathrm{BS}}$, where
\begin{align}\label{thet_aphi_BS}
\Theta_{\mathrm{BS}} = \pi - \Theta_{\mathrm{i}}\quad\text{and}\quad
\Phi_{\mathrm{BS}} = \pi \qq{,}
\end{align}
respectively.

The third step of the FHM is now complete, and the back-scattered directivity $D_{\mathrm{BS}}$ of the sphere in $S$ can be calculated by substituting Eqns.~(\ref{tot_scat_pow_simplified_lab}) and~(\ref{energy_bs_lab}) into Eqn.~(\ref{directivity_general}) such that
\begin{align}\label{directivity_bs}
   D_{\mathrm{BS}} = D(\Theta_{\mathrm{BS}}, \Phi_{\mathrm{BS}}) \qq{.}
\end{align}

\section{Relativistic Kerker condition}\label{sec:rel_kerker}
\subsection{Visualizing the directivity}

The final theoretical result of our work has been formulated in Eqn.~(\ref{directivity_bs}), which expresses the contribution of the back-scattered energy compared to the average scattered energy for a given scenario.
While Eqn.~(\ref{directivity_general}) is more general, for the sake of this discussion, we choose to investigate a possible suppression of the back-scattering (that is, the first Kerker condition) \cite{dubois2018kerker, zhang2021realizing, xiong2022polarization}. 

As an example, we consider a lossless dielectric sphere and parametrize the multipolar response using what are known as Mie angles~\cite{christoffel1957light, hulst1981light} (cf. App.~\ref{Mie_angles_definition}).
Each Mie angle is bounded between $-\sfrac{\pi}{2}$ and $\sfrac{\pi}{2}$, and we consider them to be non-dispersive.
A value of zero corresponds to a resonance of the respective Mie coefficient.

Since the objective function describing $D_{\mathrm{BS}}$ does not have a clear analytical solution, we implement numerical routines to identify properties for the sphere which minimize the back-scattering. More specifically, we wish to seek the optimized combination of Mie angles such that the back-scattering is minimized. 
Moreover, we carry out the optimization when $\beta$ and $\Theta_{\mathrm{i}}$ are fixed values; as an example we consider $\beta=0.2$ and $\Theta_{\mathrm{i}}=\frac{\pi}{4}$.

\begin{figure*}[!ht]
\centering
\includegraphics{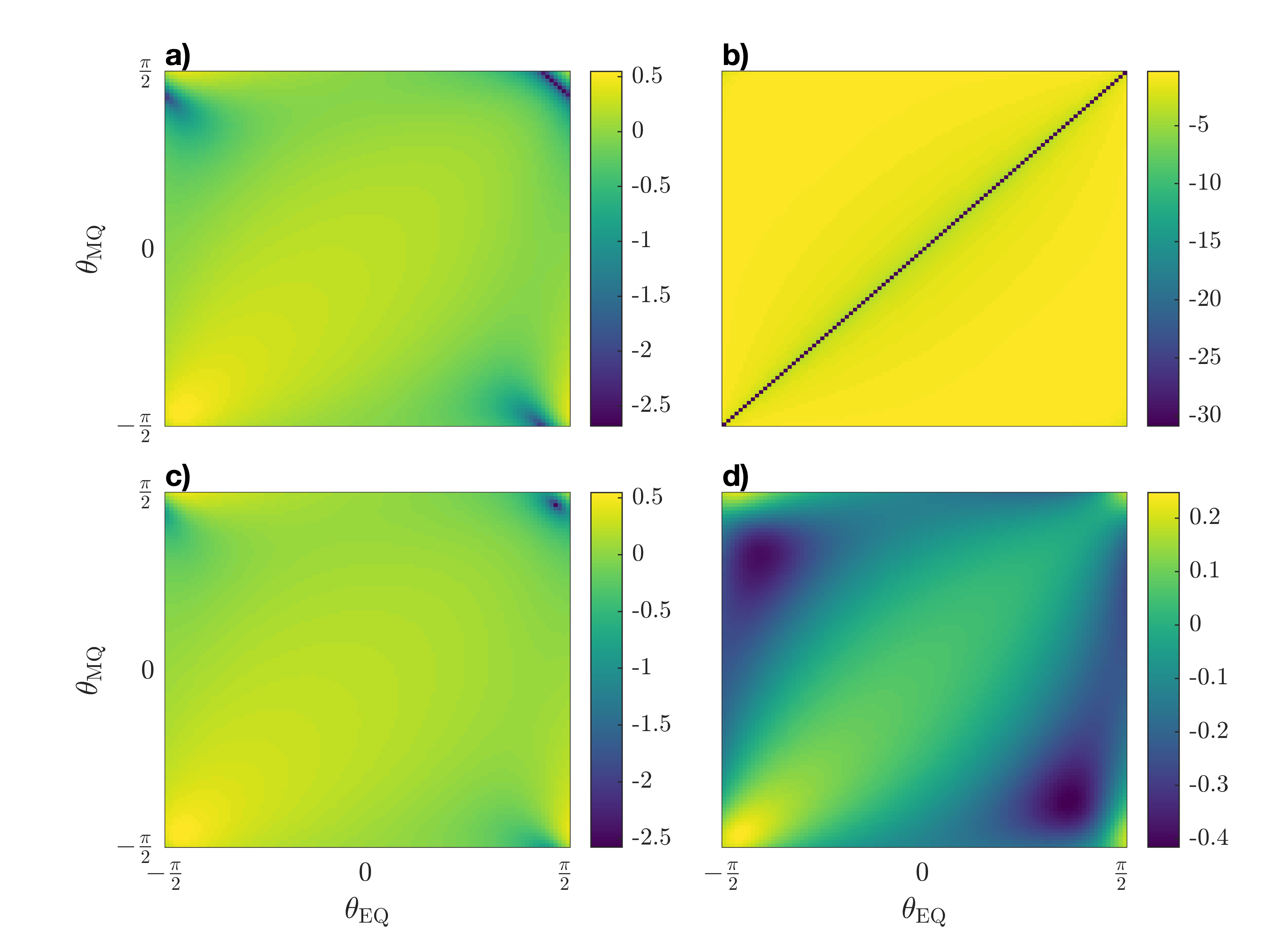}
\caption{\raggedright Figures a) and b) represent the $\lambda_{\mathrm{s}}=+1$ and $\lambda_{\mathrm{s}}=-1$ components of $\mathrm{log}_{10}D_{\mathrm{BS}}$, respectively when $\theta_{\mathrm{ED}}=\theta_{\mathrm{MD}}=\frac{\pi}{3}$. Note in b) the dark diagonal line, which means that the back-scattering vanishes when $\theta_{\mathrm{EQ}}=\theta_{\mathrm{MQ}}$. This corresponds to the case where the sphere is a dual scatterer. Since the incident helicity is given by $\lambda_{\mathrm{i}}=+1$, these vanishing components are to be expected. Figure~c) shows the total directivity $\mathrm{log}_{10}D_{\mathrm{BS}}$ when combining Figs.~a) and b), while Fig.~d) shows for comparison $\mathrm{log}_{10}D_{\mathrm{BS}}$ when $\theta_\mathrm{ED}=\frac{\pi}{9}$ and $\theta_\mathrm{MD}=-\frac{\pi}{4}$. In all cases we have that $\beta=0.2$ and $\Theta_{\mathrm{i}}=\frac{\pi}{4}$.}
\label{direc_c_combined}
\end{figure*}

Before doing this, it makes sense to visualize how $D_{\mathrm{BS}}$ varies with respect to some chosen Mie angles.
For this purpose, we sweep across the possible electric quadrupole ($\theta_{\text{EQ}}$) and magnetic quadrupole ($\theta_{\text{MQ}}$) angles while fixing the dipole angles $\theta_{\text{ED}}$ and $\theta_{\text{MD}}$.

We visualize the $\lambda_{\mathrm{s}}=+1$ and $\lambda_{\mathrm{s}}=-1$ components of $\mathrm{log}_{10}D_{\mathrm{BS}}$ with $\theta_{\mathrm{ED}} = \theta_{\mathrm{MD}} = \pi/3$ in Fig.~\ref{direc_c_combined}~a) and Fig.~\ref{direc_c_combined}~b), respectively, followed by the corresponding total directivity in Fig.~\ref{direc_c_combined}~c). In all cases, the helicity of the incident field is $\lambda_{\mathrm{i}}=1$.
One observes in Fig.~\ref{direc_c_combined}~b) that the diagonal representing $\theta_{\mathrm{EQ}} = \theta_{\mathrm{MQ}}$, that is, when the sphere is a dual scatterer, displays values of below $-30$, implying that the back-scattering is zero at these points.
This is expected since the incoming helicity is given by $\lambda_{\mathrm{i}}=+1$ and the scattered helicity in the dual case must remain the same, so we are only left with non-zero values when $\lambda_{\mathrm{s}}=+1$. For comparison, Fig.~\ref{direc_c_combined}~d) shows the total directivity where $\theta_{\mathrm{ED}}=\pi/9$ and $\theta_{\mathrm{MD}}=-\pi/4$.

\subsection{Numerically minimizing the back-scattering}

The fact that Fig.~\ref{direc_c_combined}~b) provides a physically known result is a justification of the numerical implementation and allows us to proceed in minimizing the directivity.
A further verification is the fact that the directivity is independent of the incident helicity $\lambda_{\mathrm{i}}$.
This is expected since the system is both rotationally and mirror-symmetric about the $x$-axis.
From the mirror symmetry, $\lambda_{\mathrm{i}}$ would flip sign~\cite{lamprianidis2022directional}.
However, when rotated to the same position, $\lambda_{\mathrm{i}}$ would preserve its sign.
These situations both describe the same physical scenario, that is, a system where the sphere moves in the opposite direction with $\mathbf{v}\to-v\hat{\mathbf{z}}$ and $\Theta_{\mathrm{i}}\to\pi+\Theta_{\mathrm{i}}$.
Therefore, the scattered energy remains unchanged.

To minimize the back-scattering with respect to the Mie angles, we implement the directivity calculation using the Julia programming language~\cite{bezanson2017julia} and leverage the automatic differentiation capabilities included in the modelling toolkit JuMP~\cite{dunning2017modeling} for gradient-based optimization.
This enables us to efficiently take derivatives of $D_{\mathrm{BS}}$ with respect to Mie angles up to arbitrary order.
We then formulate our optimization problem as the minimization of $D_{\mathrm{BS}}$ using the interior-point optimizer IPOPT~\cite{wachter2009short}.

Using this method, we find higher order combinations of Mie angles which yield minima below our defined cut-off point of $D_{\mathrm{C}}=10^{-3}$, that is, the value below which we consider the back-scattering to be negligible.
We consider this value appropriate, since it corresponds to a back-scattered energy which contributes a mere \SI{0.1}{\percent} to the average scattered energy.
For a single optimization run, we randomly initialize a set of Mie angles between $\qty(-\sfrac{\pi}{2}, \sfrac{\pi}{2})$ and minimize $D_{\mathrm{BS}}$ with respect to these angles.
Owing to quasi-analytical gradients, the optimization quickly converges to high-quality minima one order of magnitude lower than $D_{\mathrm{C}}$.
Finding a single set of Mie angles up to octupolar order takes less than a second on average (measured over 100 optimization runs on Intel Xeon Platinum 8368 CPU @ \SI{2.4}{\giga\hertz}).
A selection of possible combinations up to octupolar order is given in Table~\ref{table:table_combinations}.
\begin{table}[h]
\bigskip
\centering
\caption{Examples of optimized Mie angles for minimum back-scattering up to octupolar order.}
\begin{tabular}{@{}rrrrrrr@{}}
\toprule
$\theta_{\text{ED}}$ & $\theta_{\text{MD}}$ & $\theta_{\text{EQ}}$ & $\theta_{\text{MQ}}$ & $\theta_{\text{EO}}$ & $\theta_{\text{MO}}$ & $D_{\mathrm{BS}}\times10^{-4}$ \\
\midrule
$-0.18$ & $1.21$ & $1.38$ & $1.23$ & $1.54$ & $1.55$ & $1.57$ \\
$-1.39$ & $-1.32$ & $-1.12$ & $-1.44$ & $-1.51$ & $-1.57$ & $1.95$ \\
$-1.32$ & $-1.32$ & $-1.21$ & $-1.21$ & $-1.53$ & $-1.53$ & $2.02$ \\
\bottomrule
\end{tabular}
\label{table:table_combinations}
\end{table}

These minimized values of $D_{\mathrm{BS}}$ are all smaller than $D_{\mathrm{T}}=10^{-3}$ and thus satisfy our cut-off criterion, providing evidence of the existence of the first Kerker condition in the relativistic regime.
The most pronounced minimum located ($D_{\mathrm{BS}}=1.57\times 10^{-4}$) describes a case where the back-scattered energy contributes a negligible \SI{0.016}{\percent} to the average scattered energy.
It must be emphasized that there exist many combinations of Mie angles that fulfil this condition, making those in Table~\ref{table:table_combinations} just few of many.

\section{Conclusion}\label{sec:conclusion}

The main goal of this paper was to demonstrate the utility of expressing incident and scattered fields in the helicity basis for the case of a sphere moving at a relativistic speed.
In doing this, we were able to transform fewer variables (namely the helicity) as opposed to the same problem in the parity basis, thus greatly simplifying calculations.
Moreover, we obtained an expression for the back-scattering amplitude of the scattered field observed from an external lab frame in the form of the directivity of the sphere.
Finally, the directivity was minimized with respect to Mie angles, providing evidence for the existence of the first Kerker condition in the relativistic regime.

Opportunities for future work are plentiful.
Since our implementation can be differentiated with respect to all continuous parameters of the considered system, it is, in principle, possible to optimize not only for the Mie angles but also for parameters beyond the scope of this work, such as incident angle and velocity.
Furthermore, the current work could be extended to analyze a composition of particles describing a metasurface as opposed to just a single particle.
In this case, a cluster T-matrix as described in~\citet{mishchenko2002scattering} would need to be implemented.
The motivation for considering a surface of particles links to the future applications of light sails proposed by the Breakthrough Starshot Initiative.

Of course, for this to be done, the current model would have to be refined to consider the opposite case of maximum back-scattering, resulting in the ideal case of maximum momentum transfer to the sail~\cite{tung2022low}.

\section*{Data availability}\label{data_availability}
The code used to produce the results in this manuscript can be accessed via the following link: \url{https://github.com/tfp-photonics/Jorkle.jl}. 

\begin{acknowledgments}
M.R.W., A.G.L., and C.R. acknowledge support from the Max Planck School of Photonics, which is supported by BMBF, Max Planck Society, and the Fraunhofer Society. M.R.W. and A.G.L. also acknowledge support from the Karlsruhe School of Optics and Photonics (KSOP).
Y.A. and C.R. acknowledge support by the German Research Foundation within the Excellence Cluster 3D Matter Made to Order (EXC 2082/1 under project number - 390761711) and by the Carl Zeiss Foundation.
The optimizations in Section~\ref{sec:rel_kerker} were carried out on the HoreKa supercomputer funded by the Ministry of Science, Research and the Arts Baden-Württemberg and by the Federal Ministry of Education and Research.
\end{acknowledgments}

\appendix

\section{Lorentz boost of helical plane waves}\label{LB_of_plane_waves}

As stated in~\citet{garner2017time}, the Lorentz boost of the electric field $\mathbf{E}(\mathbf{r},t)$ is given by
\begin{align}\label{EF_lorentz_boost}
    \mathbf{E}'(\mathbf{r}',t') &= \gamma\left[\mathbf{E}(\mathbf{r},t)+ v\hat{\mathbf{v}}\times\mathbf{B}(\mathbf{r},t)\right]\nonumber\\
    &+(1-\gamma)\left[\hat{\mathbf{v}}\cdot\mathbf{E}(\mathbf{r},t)\right]\hat{\mathbf{v}} \qq{,}
\end{align}
where $\mathbf{B}(\mathbf{r},t)$ is the corresponding magnetic field and $\gamma=1/\sqrt{1-\beta^2}$ with $\beta=v/c$ being the ratio of the boosting speed to the speed of light and boosting takes place along the direction $\hat{\mathbf{v}}$. For boosting along $+\hat{\mathbf{z}}$, the coordinates of the primed (boosted) and unprimed coordinate systems are related with the following formulas:
\begin{eqnarray}
x'&=&x\label{eq:coordtrans1}\\
    y'&=&y\label{eq:coordtrans2}\\
    z'&=&\gamma(z-\beta ct) \label{eq:coordtrans3}\\
    t'&=&\gamma(t-\beta z/c) \qq{.} \label{eq:coordtrans4}
\end{eqnarray}
Here, we want to consider the boost of a monochromatic plane wave of well-defined helicity $\lambda$. Specifically, its electric field is given by $\mathbf{E}(\mathbf{r},t)=\hat{\mathbf{e}}_{\lambda}(\hat{\mathbf{k}})\exp\{{\mathrm{i}\omega[(\hat{\mathbf{k}}\cdot\mathbf{r}/c) - t}]\}$ (see main text). By making use of the above coordinate transformations we can get the following transformation of the exponent: $\{{\mathrm{i}\omega[(\hat{\mathbf{k}}\cdot\mathbf{r}/c) - t}]\}=\{{\mathrm{i}\omega'[(\hat{\mathbf{k}}'\cdot\mathbf{r}'/c) - t'}]\}$, with $\omega',\hat{\mathbf{k}}'$ being given by Eqs.~(\ref{freq_boost},\ref{LB_k}) of the main text. Thus, we have transformed the scalar part of the fields, which gave us the transformed frequency and direction of propagation of the boosted plane wave, $\omega',\hat{\mathbf{k}}'$, respectively.

Next, we need to transform the polarization vector. For this we need to take into account that our considered plane wave, being an eigenstate of the helicity operator $\frac{\nabla\times}{k}$ with well-defined helicity $\lambda$~\cite{fernandez2021total}, has the following property:
\begin{align}\label{MEQ}
    \nabla\times\mathbf{E}(\mathbf{r},t) 
    &= \lambda k\mathbf{E}(\mathbf{r},t) \qq{,}
\end{align}
and, therefore, we get the following for its corresponding magnetic field from Maxwell's equations:
\begin{align}\label{B=const_times_E}
   \mathbf{B}(\mathbf{r},t) &= \frac{\lambda}{\mathrm{i}c}\mathbf{E}(\mathbf{r},t) \qq{.}
\end{align}
Therefore, by substituting the right-hand-side of Eqn.~(\ref{B=const_times_E}) into Eqn.~(\ref{EF_lorentz_boost}), we find that
\begin{align}\label{EF_inc_scatterer_frame}
    \mathbf{E}'(\mathbf{r}',t') &= \gamma\left[\mathbf{E}(\mathbf{r},t)+\frac{\lambda v}{\mathrm{i}c}\hat{\mathbf{z}}\times\mathbf{E}(\mathbf{r},t)\right]\nonumber\\
    &+(1-\gamma)\left[\hat{\mathbf{z}}\cdot\mathbf{E}(\mathbf{r},t)\right]\hat{\mathbf{z}} \qq{.}
\end{align}
What remains then, is to project the boosted helical polarization vector $\hat{\mathbf{e}}_{\lambda}(\hat{\mathbf{k}})$ onto the boosted polarization basis $\hat{\mathbf{e}}_{\lambda'}(\hat{\mathbf{k}}')$. That is to say, we need to find the coefficients $\mathcal{E}_{\lambda\lambda'}(\beta, \hat{\mathbf{k}})$ in the expansion bellow:
\begin{align}\label{Lorentz_boost_polarisation}
&\gamma\left[\hat{\mathbf{e}}_{\lambda}(\hat{\mathbf{k}}) + \frac{\lambda v}{\mathrm{i}c}\hat{\mathbf{z}}\times\hat{\mathbf{e}}_{\lambda}(\hat{\mathbf{k}})\right]\nonumber\\
&+(1-\gamma)\left[\hat{\mathbf{z}}\cdot\hat{\mathbf{e}}_{\lambda}(\hat{\mathbf{k}})\right]\hat{\mathbf{z}}\nonumber\\
&= \sum_{\lambda'}\mathcal{E}_{\lambda\lambda'}(\beta, \hat{\mathbf{k}})\hat{\mathbf{e}}_{\lambda'}(\hat{\mathbf{k}}') \qq{.}
\end{align}
By making use of the following orthogonality relation~\cite{lamprianidis2022directional}:
\begin{equation}\label{orthog_rel}
    \hat{\mathbf{e}}_{\lambda'}(\mathbf{k}')\cdot\hat{\mathbf{e}}_{-\lambda'_0}(\mathbf{k}')=-\delta_{\lambda'\lambda'_0} \qq{,}
\end{equation}
we readily get after some algebra that $\mathcal{E}_{\lambda\lambda'}(\beta, \hat{\mathbf{k}})=\delta_{\lambda\lambda'}\mathcal{C}_{\lambda}(\beta, \theta)$, with $\theta$ being the polar angle of the propagation direction $\hat{\mathbf{k}}$ and $\mathcal{C}_{\lambda}(\beta, \theta)$ being given by:
\begin{align}\label{simplified_C_app}
    \mathcal{C}_{\lambda}(\beta, \theta)
    &=
    \gamma\left[1-\beta\cos\theta\right] \qq{.}
\end{align}
The same expression calculated using the parity basis is given by Eqn.~(27) in~\citet{de2000electromagnetic}. Note that the helicity $\lambda$ of massless particles (and hence, elecromagnetic fields) is invariant under Lorentz boosts~\cite{fernandez2015helicity}. Finally, summing up all the above, we get the Lorentz boost transformation given by Eq.~(\ref{projection}) in the main text.

\section{Vector spherical harmonics of well-defined helicity}\label{VSH_definitions}

We  begin by using the following definition of the spherical harmonics:
\begin{eqnarray}
\mathrm{Y}_\ell^m(\theta,\phi)&\triangleq&\Omega_{\ell m}P_\ell^m(\mathrm{cos}\theta)\mathrm{e}^{\mathrm{i}m\phi} \qq{,}
\end{eqnarray}
where $P_\ell^m (\mathrm{cos}\theta)$ is the associated Legendre function of the 1st kind, with $\Omega_{\ell m}\triangleq \mathrm{i}^m \sqrt{\frac{(2\ell+1)(\ell-m)!}{4\pi\ell(\ell+1)(\ell+m)!}}$ being the corresponding normalization factor.

Then, the VSHs of well-defined parity, $\mathbf{M}_{\ell m k}^{(j)}$ and $\mathbf{N}_{\ell m k}^{(j)}$, are defined as follows~\cite{feshbach}:
\begin{align}
\mathbf{M}_{\ell m k}^{(j)} \left( \mathbf{r} \right) 
&\triangleq
\nabla \times\left[\mathbf{r}z^{(j)}_{\mathrm{M},\ell}(kr )\mathrm{Y}_\ell^m(\theta,\phi)\right]\nonumber\\
&=
\mathrm{i}z_{\mathrm{M},\ell}^{(j)}(kr)\mathbf{m}_{\ell m}(\hat{\mathbf{r}}),  \label{sfairM}\\
\mathbf{N}_{\ell m k}^{(j)} \left(  \mathbf{r} \right) 
&\triangleq
\frac{1}{k}\nabla \times\mathbf{M}_{\ell m k}^{(j)} \left( \mathbf{r} \right)\nonumber\\
&=
\hat{\mathbf{r}}\frac{\ell(\ell+1)}{\mathrm{k_0r}}z_{\mathrm{M},\ell}^{(j)}(kr) \mathrm{Y}_\ell^m(\theta,\phi) \nonumber\\
&+
z_{\mathrm{N},\ell}^{(j)}(kr)\mathbf{n}_{\ell m}(\hat{\mathbf{r}}) \qq{,} \label{sfairN}
\end{align}
where
\begin{eqnarray}
\mathbf{m}_{\ell m}(\hat{\mathbf{r}}) &=&\Omega_{\ell m}\left[\hat{\theta}\tau_{\ell m}(\theta) + \mathrm{i}\hat{\phi}\tau_{\ell m}'(\theta) \right]\mathrm{e}^{\mathrm{i}m\phi},
\\
\mathbf{n}_{\ell m}(\hat{\mathbf{r}}) &=&  \Omega_{\ell m}\left[\hat{\theta}\tau_{\ell m}'(\theta) +  \mathrm{i}\hat{\phi}\tau_{\ell m}(\theta) \right]\mathrm{e}^{\mathrm{i}m\phi} \qq{.}
\end{eqnarray}
The index $\ell$ stands for the angular momentum quantum number that takes the values 1,2,$\dots$ and corresponds to dipoles, quadrupoles, etc., and the index $m$ stands for the angular momentum along the $z$-axis which takes the values $-\ell,...,-2,-1,0,1,2,...,\ell$. The superscript $j$ refers to the corresponding Bessel ($j=1$) and Hankel ($j=3$) functions, $z_{\mathrm{M},\ell}^{(j)}(kr)$, of the first kind. The functions $z_{\mathrm{N},\ell}^{(j)}(kr)\triangleq\frac{1}{kr}\frac{\mathrm{d}}{\mathrm{d}(kr)}[krz_{\mathrm{M},\ell}^{(j)}(kr)]$ are the corresponding Riccati functions and $\tau_{\ell m}(\theta) \triangleq m\frac{P_\ell^m(\mathrm{cos}\theta)}{\mathrm{sin}\theta}$ and $\tau_{\ell m}'(\theta) \triangleq
\frac{\mathrm{d}P_\ell^m(\mathrm{cos}\theta)}{\mathrm{d}\theta}$ are the generalized Legendre functions.

The VSHs $\mathbf{\Lambda}^{(j)}_{\lambda,\ell m k}(\mathbf{r})$ of well-defined helicity $\lambda=\pm1$ are defined with respect to the VSHs of well-defined parity according to the formula:
\begin{align}
\mathbf{\Lambda}^{(j)}_{\lambda,\ell m k}(\mathbf{r})
&=
\frac{\mathbf{M}^{(j)}_{\ell m k}(\mathbf{r})+\lambda\mathbf{N}^{(j)}_{\ell m k}(\mathbf{r})}{\sqrt{2}}\\
&=
\frac{\lambda}{\sqrt{2}}\frac{\ell(\ell+1)}{kr}z_{\mathrm{M},\ell}^{(j)}(kr) \mathrm{Y}_\ell^m(\theta,\phi) \hspace{3pt}\hat{\mathbf{r}}\nonumber\\
&+
\sum_{\lambda'=\pm1}\Big[\frac{\mathrm{i}z^{(j)}_{\mathrm{M},\ell}(k r)+\lambda\lambda'z^{(j)}_{\mathrm{N},\ell}(k r)}{2}\nonumber\\
&\cdot
\hspace{0.2cm}\mathbf{f}_{\lambda',\ell m}(\hat{\mathbf{r}})\Big] \qq{,}
\end{align}
where we have defined:
\begin{align}\mathbf{f}_{\lambda,\ell m}(\hat{\mathbf{r}})
&=
\frac{\mathbf{m}_{\ell m}(\hat{\mathbf{r}})+\lambda\mathbf{n}_{\ell m}(\hat{\mathbf{r}})}{\sqrt{2}}\nonumber\\
&=
\Omega_{\ell m}\tau^{(\lambda)}_{\ell m}(\theta )\mathrm{e}^{\mathrm{i}m\phi}\hspace{3pt}\hat{\mathbf{e}}_\lambda(\hat{\mathbf{r}})
\end{align}
and
\begin{eqnarray}\tau^{(\lambda)}_{\ell m}(\theta )&=&-\tau'_{\ell m}(\theta)-\lambda\tau_{\ell m}(\theta ) \qq{,}
\end{eqnarray}
which has the property $\Omega_{-\ell m}\tau^{(\lambda)}_{-\ell m}(\theta )=\Omega_{\ell m}\tau^{(-\lambda)}_{\ell m}(\theta )=(-1)^{\ell+m+1}\Omega_{\ell m}\tau^{(\lambda)}_{\ell m}(\pi-\theta )$.

One can show that the functions $\mathbf{\Lambda}^{(j)}_{\lambda,\ell m k}$ have the property~\cite{fernandez2021total}:
\begin{eqnarray}
    \frac{\nabla\times}{k}\mathbf{\Lambda}^{(j)}_{\lambda,\ell m k}&=&\lambda\mathbf{\Lambda}^{(j)}_{\lambda,\ell m k} \qq{,}
\end{eqnarray}
that is, $\mathbf{\Lambda}^{(j)}_{\lambda,\ell m k}$ is an eigenstate of the helicity operator $\frac{\nabla\times}{k}$ with eigenvalue $\lambda$. For the functions $\mathbf{f}_{\lambda,\ell m}$, there exists the orthogonality property:
\begin{align}\label{f_orthogonality}
    \int_0^{2\pi}\mathrm{d}\phi\int_0^{\pi}&\mathrm{sin}\theta\mathrm{d}\theta\nonumber\\
    &\mathbf{f}_{\lambda,\ell m}(\hat{\mathbf{k}})\cdot\left[\mathbf{f}_{\lambda',\ell'm'}(\hat{\mathbf{k}})\right]^*\nonumber\\
    &=
    \delta_{\lambda\lambda'}\delta_{\ell' m'}\delta_{\ell' m'} \qq{.}
\end{align}
Moreover, if we employ the large argument property of the Hankel functions: 
\begin{align}\label{Hankel_ff}
z_{\alpha, \ell}^{(3)}(x) \xrightarrow{x>>1}\begin{cases}
        \frac{\mathrm{e}^{\mathrm{i}x}}{x} (-\mathrm{i})^{\ell} & \text{for}\quad \alpha=\mathrm{N}\\
        \frac{\mathrm{e}^{\mathrm{i}x}}{x} (-\mathrm{i})^{\ell+1} & \text{for}\quad \alpha=\mathrm{M}\qq{,}
    \end{cases}
\end{align}
and also reject the $\mathrm{O}(\sfrac{1}{\mathrm{r_0^2}})$ radial term as negligible, for the radiating helical VSHs we can get the following asymptotic form in the far field:
\begin{eqnarray}
	\left[\mathbf{\Lambda}_{\lambda,\ell m k}^{(3)} \left( \mathbf{r} \right)\right]^{\mathrm{ff}}&=&(-\mathrm{i})^{\ell}\hspace{3pt}\mathbf{f}_{\lambda,\ell m}(\hat{\mathbf{r}})\hspace{3pt}\frac{\mathrm{e}^{\mathrm{i}kr}}{kr} \qq{.}
\end{eqnarray}

\section{T-matrix in the helicity basis for a sphere}\label{T-matrix_helicity}

In the parity basis, the T-matrix is given by
\begin{equation}\label{t_matrix_dps}
\mathbf{T}=\begin{pmatrix}
\mathbf{T}_{\text{NN}} & \mathbf{T}_{\text{MN}}\\
\mathbf{T}_{\text{NM}} & \mathbf{T}_{\text{MM}}
\end{pmatrix} \qq{,}
\end{equation}
where each element of Eqn.~(\ref{t_matrix_dps}) is a diagonal $\ell_{\text{max}}(2+\ell_{\text{max}})\times\ell_{\text{max}}(2+\ell_{\text{max}})$ matrix and $\ell_{\text{max}}$ is the maximum multipolar excitation order of the sphere.

This can be transformed to the T-matrix $\mathbf{T}^{\text{H}}$ in the helicity basis by using~\cite{rahimzadegan2018core}
\begin{equation}\label{t_matrix_transform}
\mathbf{T}^{\text{H}} = \mathbf{P}^{-1}\mathbf{T}\mathbf{P} \qq{,}
\end{equation}
where, as can be seen from Eqn.~(\ref{polarisation_unit_vec}), $\mathbf{P}$ is given by 
\begin{equation}\label{P}
\mathbf{P}=\frac{1}{\sqrt{2}}\begin{pmatrix}
1 & 1\\
1 & -1
\end{pmatrix} \qq{.}
\end{equation}

In the case of a dielectric sphere, we have that $\mathbf{T}_{\text{MN}} = \mathbf{T}_{\text{NM}} = 0$, so Eqn.~(\ref{t_matrix_transform}) reduces to
\begin{align}\label{T_H_expanded_2}
\mathbf{T}^{\text{H}}
&=
\begin{pmatrix}
\mathbf{T}_{++} & \mathbf{T}_{+-}\\
\mathbf{T}_{-+} & \mathbf{T}_{--}
\end{pmatrix}\nonumber\\
&=
\frac{1}{2}\begin{bmatrix}
(\mathbf{T}_{\text{NN}}+\mathbf{T}_{\text{MM}}) & (\mathbf{T}_{\text{NN}}-\mathbf{T}_{\text{MM}})\\
(\mathbf{T}_{\text{NN}}-\mathbf{T}_{\text{MM}}) & (\mathbf{T}_{\text{NN}}+\mathbf{T}_{\text{MM}})
\end{bmatrix} \qq{,}
\end{align}
where 
\begin{equation}\label{T_NN}
\mathbf{T}_{\text{NN}}=\begin{pmatrix}
a_{1} & \dots & 0\\
\vdots & \ddots & \vdots\\
0 & \dots & a_{\ell_{\text{max}}}
\end{pmatrix} \qq{,}
\end{equation}
\begin{equation}\label{T_MM}
\mathbf{T}_{\text{MM}}=\begin{pmatrix}
b_{1} & \dots & 0\\
\vdots & \ddots & \vdots\\
0 & \dots & b_{\ell_{\text{max}}}
\end{pmatrix} \qq{,}
\end{equation}
and the values $a_{\ell}$ and $b_{\ell}$ are respectively the electric and magnetic Mie coefficients defined in App.~\ref{Mie_angles_definition}.

The components of the T-matrix $\mathbf{T}^{\mathrm{H}}$ are given by $\mathrm{T}_{\lambda_{\mathrm{s}},\lambda_{\mathrm{i}},\ell}$, and relate to the entries in $\mathbf{T}^{\mathrm{H}}$ corresponding to the $\ell$'th multipolar order, along with an incident helicity $\lambda_{\mathrm{i}}$ and scattered helicity $\lambda_{\mathrm{s}}$. That is, 

\begin{align}\label{T_mat_components}
\mathrm{T}_{\lambda_{\mathrm{s}},\lambda_{\mathrm{i}},\ell} 
&=
a_{\ell} + \lambda_{\mathrm{i}}\lambda_{\mathrm{s}}b_{\ell}.
\end{align}

\section{Computation-friendly expansion of Eqn.~(\ref{ang_pow_integral_sphere_frame})}\label{expansion_U}

By substituting Eqn.~(\ref{E_sca_VSH_ff}) into the Eqn.~(\ref{ang_pow_integral_sphere_frame}), and using the orthogonality relation given by Eqn.~(\ref{f_orthogonality}), we can express the  angular energy density $U'(\theta', \phi')$ in the following form that is suitable for efficient numerical evaluation:
\begin{align}\label{ang_pow_integral_sphere_frame_processed}
U'(\theta', \phi') &= 
\sum_{\lambda'}\frac{4\pi c^{2}}{\eta_{0}}\int^{\infty}_{0^{+}}\mathrm{d}\omega'
    \frac{1}{(\omega')^2}\nonumber\\
    &\cdot
    \left|\sum_{\ell'm'}\mathcal{B}_{\lambda'\ell'm'}(\omega')(-\mathrm{i})^{\ell'}\Omega_{\ell'm'}\tau^{(\lambda')}_{\ell'm'}(\theta')\mathrm{e}^{\mathrm{i} \ell'\phi'}\right|^2\nonumber\\
    &=
    \sum_{\lambda'}\sum_{\ell'm'}\sum_{\bar{\ell}'\bar{m'}}Q_{\lambda'\ell'm'}^{\bar{\ell}'\bar{m'}}(\theta',\phi')\nonumber\\
    &\cdot
    \sum_{\lambda_0\bar{\lambda}_0}J_{\lambda_0\ell'm'}^{\bar{\lambda}_0\bar{\ell}'\bar{m'}}\mathrm{T}_{\lambda'\lambda_0,\ell'}\mathrm{T}_{\lambda'\bar{\lambda}_0,\bar{\ell}'}^* \qq{,}
\end{align}
where the T-matrix elements $\mathrm{T}_{\lambda'\lambda_0,\ell'}$ are defined in App.~\ref{T-matrix_helicity}. For the latter equation, we have assumed a non-dispersive T-matrix and have also defined the integral:
\begin{align}\label{Jlm}
J_{\lambda_0\ell'm'}^{\bar{\lambda}_0\bar{\ell}'\bar{m'}}&=&\int^{\infty}_{0^{+}}\mathrm{d}\omega'
\frac{\mathcal{A}_{\lambda_0\ell'm'}(\omega')\mathcal{A}^{*}_{\bar{\lambda}_0\bar{\ell'}\bar{m'}}(\omega')}{(\omega')^2} \qq{,}
\end{align}
and the quantity:
\begin{align}\label{Qlm}
   Q_{\lambda'\ell'm'}^{\bar{\ell}'\bar{m'}}(\theta',\phi')
   &= 
   \frac{4\pi c^{2}}{\eta_{0}}(-\mathrm{i})^{\ell'-\bar{\ell}'}\Omega_{\ell'm'}\Omega^*_{\bar{\ell}'\bar{m'}}\nonumber\\
   &\cdot
   \tau^{(\lambda')}_{\ell'm'}(\theta')\tau^{(\lambda')*}_{\bar{\ell}'\bar{m'}}(\theta')\mathrm{e}^{\mathrm{i} (m'-\bar{m'})\phi'} \qq{,}
\end{align}
where $^*$ denotes the complex conjugate.

Furthermore, we can define express the total scattered energy $W_{\mathrm{tot}}$ given by Eqn.~(\ref{tot_scat_pow_simplified_lab}) as follows:

\begin{align}\label{W_computational}
W_\mathrm{tot}
&=
\int_0^\pi \mathrm{d}\theta \int_0^{2\pi}\mathrm{d}\phi\sin\theta U(\theta, \phi)\nonumber\\
    &=
    \sum_{\lambda'}\int^{\infty}_{0^{+}}\mathrm{d}\omega'
    \frac{1}{(\omega')^2}\sum_{\ell' m';\mathrm{min}\{|m'|,1\}\leq\ell\leq\ell'}\nonumber\\
    &
    \hspace{0.5cm}\Re\left\{\mathcal{B}_{\omega'\lambda'\ell' m'}\mathcal{B}^*_{\omega'\lambda'\ell m'}I_{\lambda'\ell' m'}^{\ell}\right\}\nonumber\\
    &=
    \sum_{\lambda'}\sum_{\ell' m';\mathrm{min}\{|m'|,1\}\leq\ell\leq\ell'}\sum_{\lambda_0\bar{\lambda}_0}\nonumber\\
    &\Re\left\{I_{\lambda'\ell' m'}^{\ell}J_{\lambda_0m'\ell'}^{\bar{\lambda}_0m'\ell}\mathrm{T}_{\lambda'\lambda_0,\ell'}\mathrm{T}_{\lambda'\bar{\lambda}_0,\ell}^*\right\}
\end{align}
with
\begin{align}\label{I_integral}
I_{\lambda'\ell' m'}^{\ell}
&=
2^{4-\delta_{\ell\ell'}}\frac{\pi^2 c^{2}}{\eta_{0}}  (-\mathrm{i})^{\ell'-\ell}\Omega_{\ell' m'}\Omega_{\ell m'}^*\nonumber\\
&\int_0^\pi
\mathrm{d}\theta \sin\theta \left\{\gamma\left[1+\beta\,\theta'(\beta, \theta)\right]\right\}^{3}\nonumber\\
&\times
\tau^{(\lambda')}_{\ell' m'}\left[\theta'(\beta, \theta)\right]
\tau^{(\lambda')*}_{\ell m'}\left[\theta'(\beta, \theta)\right]\qq{,}
\end{align}
where the expression for $\theta'(\beta, \theta)$ is given by Eqn.~\eqref{theta_boost}.

\begin{figure}[ht]
\vspace{0.1cm}
\centering
\includegraphics{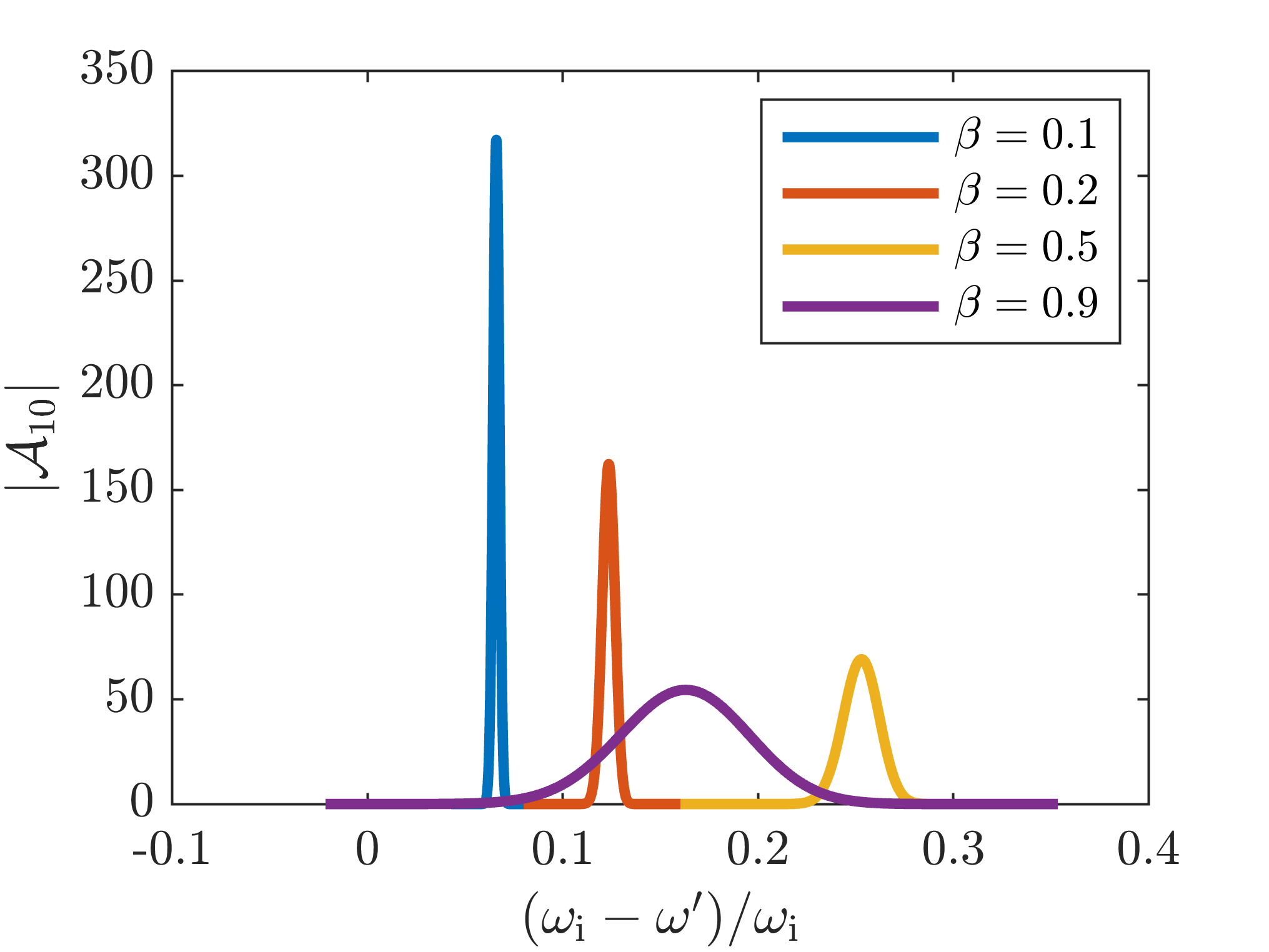}
\caption{\raggedright The absolute value of the dipole expansion coefficients $\mathcal{A}_{10}$ ($\ell'=1, m'=0$) given by Eqn.~(\ref{A_simplified}) as a function of the normalized frequency $(\omega_{\mathrm{i}}-\omega')/\omega_{\mathrm{i}}$ when a): $\beta=0.1$ (blue line), $\beta=0.2$ (orange line), $\beta=0.5$ (yellow line) and $\beta=0.9$ (purple line). In all cases, $\Theta_{\mathrm{i}}=\frac{\pi}{4}$, and  $\lambda_{\mathrm{i}}=+1$ which, since helicity is conserved under Lorentz boosts, means that $\lambda'=\lambda_{\mathrm{i}}=+1$. Note that the width of the peak increases with speed. This is due to the increasing effect of the Doppler shift $\omega'$ of $\omega_{\mathrm{i}}$. Since the incident beam is monochromatic, $|\mathcal{A}_{10}|$ tends to a delta-distribution-like peak as the speed decreases. As the speed increases, the plane wave components of the incident beam all Doppler shift differently due to their differing angular orientations. This leads to non-zero values for $|\mathcal{A}_{10}|$ when $\omega_{\mathrm{i}}\neq\omega'$.}
\label{combined_A}
\end{figure}

Note the importance of writing Eqn.~(\ref{ang_pow_integral_sphere_frame_processed}) as nested integrals instead of a standard triple integral. Computationally speaking, we are able to determine Eqn.~(\ref{Jlm}) with a very low tolerance, while using a higher tolerance for the other integrals, thus significantly reducing computation time. The reason for this is because the expansion coefficients $\mathcal{A}_{\lambda'\ell m}(\omega')$ (and hence the integrand in Eqn.~(\ref{Jlm})) form Gaussian-like peaks centered about $\omega'$ which tend to a delta distribution as the speed of the sphere decreases (see Fig.~\ref{combined_A}). If the tolerance is too high, the numerical integration could miss this peak entirely, thus ignoring vital non-zero values.

Moreover, by separating the integrals Eqn.~(\ref{W_computational}) we are able to obtain $100\times100$ grids for the directivity (like those used to generate Fig.~\ref{direc_c_combined}) in as little as $\sim$\SI{40}{\second}. The reason for this is that, as the numerically-demanding integral $J_{\lambda_0\ell'm'}^{\bar{\lambda}_0\bar{\ell}'\bar{m'}}$ is independent of the Mie angles, we only need to compute it once for a given incident angle $\Theta_{\mathrm{i}}$ and speed parameter $\beta$. If the integrals were combined, $J_{\lambda_0\ell'm'}^{\bar{\lambda}_0\bar{\ell}'\bar{m'}}$ would be computed for each combination of Mie angles, thus significanly increasing computation time. 

\section{Mie angles for a lossless sphere}\label{Mie_angles_definition}

The electric and magnetic Mie coefficients $a_{\ell}$ and $b_{\ell}$, respectively for each multipolar order $\ell$ can be represented using Mie angles $\theta_{\text{E}\ell}$ and $\theta_{\text{M}\ell}$. In the lossless case, one can write~\cite{hulst1981light, rahimzadegan2020minimalist}
\begin{equation}\label{Mie_coeffs_NA_a}
    a_{\ell} = -\mathrm{i}\sin{\alpha_{\ell}}\exp(-\mathrm{i}\alpha_{\ell})
\end{equation}
and
\begin{equation}\label{Mie_coeffs_NA_b}
    b_{\ell} = -\mathrm{i}\sin{\beta_{\ell}}\exp(-\mathrm{i}\beta_{\ell}) \qq{,}
\end{equation}
where
\begin{equation}\label{alpha_i}
    \alpha_{\ell} = \frac{\pi}{2} - \theta_{\text{E}\ell},\quad-\frac{\pi}{2}\leq\theta_{\text{E}\ell}\leq\frac{\pi}{2}\qq{,}
\end{equation}
and
\begin{equation}\label{beta_i}
    \beta_{\ell} = \frac{\pi}{2} - \theta_{\text{M}\ell}, \quad-\frac{\pi}{2}\leq\theta_{\text{M}\ell}\leq\frac{\pi}{2} \qq{.}
\end{equation}
Note that the convention used in~\citet{rahimzadegan2020minimalist} omits the use of the minus sign in Eqns.~(\ref{Mie_coeffs_NA_a}) and~(\ref{Mie_coeffs_NA_b}). In our case, the minus sign is required for energy conservation.

\FloatBarrier
\bibliography{references}
%

\end{document}